\documentclass[fleqn,12pt,twoside]{article}
\usepackage{espcrc1}

\usepackage{graphicx,colordvi}

\def\prt{\partial}

\newcommand{\AmS}{{\protect\the\textfont2
  A\kern-.1667em\lower.5ex\hbox{M}\kern-.125emS}}

\title{Quark-mass dependence of baryon resonances}

\author{M.F.M. Lutz
\address[GSI]{Gesellschaft f\"ur Schwerionenforschung (GSI),\\
Planck Str. 1, D-64291 Darmstadt, Germany}
\address{Institut f\"ur Kernphysik, TU Darmstadt\\
D-64289 Darmstadt, Germany},
C. Garc\'\i a-Recio
\address[Granada]{Departamento de F\'\i sica
Moderna, \\Universidad de Granada, E-18071 Granada, Spain},
E.E. Kolomeitsev
\address[NBI]{The Niels Bohr Institute\\ Blegdamsvej 17, DK-2100 Copenhagen\\Denmark}
and J. Nieves
\addressmark[Granada]}

\begin{document}

\maketitle

\begin{abstract}
We study the quark-mass dependence of $J^P =
\frac12^-$ s-wave and $J^P =
\frac32^-$ d-wave baryon resonances. Parameter-free results are
obtained in terms of the
leading order chiral Lagrangian. In the 'heavy' SU(3) limit with
$m_\pi =m_K \simeq $ 500 MeV the s-wave resonances turn into bound states
forming two octets plus a singlet representations of the SU(3) group. Similarly the
d-wave resonances turn into bound states forming an octet and a decuplet in this limit.
A contrasted result is obtained in the 'light' SU(3) limit with $m_\pi
=m_K \simeq $ 140 MeV for which no resonances exist.
\end{abstract}

\section{Introduction}

The meson-baryon scattering processes are an important test for effective field theories
which aim at reproducing QCD at small energies, where the effective degrees of freedom are
hadrons rather than quarks and gluons. The task to construct a systematic effective field
theory for the meson-baryon scattering
processes in the resonance region is closely linked to the fundamental question
as to what is the 'nature' of baryon resonances. Here we support the conjecture
\cite{LK01,LK02,LWF02,LH02} that baryon resonances not belonging to the large-$N_c$ ground
states are generated dynamically by coupled-channel dynamics
\cite{Wyld,Dalitz,Ball,Rajasekaran,Wyld2,sw88}. For a comprehensive
discussion of this issue we refer to \cite{LH02}. This conjecture was the basis of
the phenomenological model \cite{LWF02}, which generated successfully non-strange
s- and d-wave resonances by coupled-channel dynamics describing a large body of pion and
photon scattering data. Before the event of the quark-model it was
already suggested by Wyld \cite{Wyld} and also by Dalitz, Wong and Rajasekaran \cite{Dalitz}
that a $t$-channel vector meson exchange model for the s-wave
meson-baryon scattering problem has the potential to dynamically
generate s-wave baryon resonances upon solving a coupled-channel Schr\"odinger equation.
In a more modern language the $t$-channel exchange was rediscovered in terms of the
Weinberg-Tomozawa (WT) interaction, the leading term of the chiral
Lagrangian that reproduces the first term of the vector meson exchange
in an appropriate Taylor expansion \cite{Wein-Tomo}. The main difference of the early attempts
from computations based on the chiral Lagrangian is the way the coupled-channel scattering equation
is regularized and renormalized. The crucial advance over the last years in this field is therefore
a significant improvement of the systematics, i.e. how to implement corrections terms into
coupled-channel dynamics \cite{LK00,LK01,LK02}.

In recent works \cite{LK01,LK02,LK00,Granada,KL03}, which will be reviewed
here, it was shown that chiral dynamics provides a parameter-free prediction for the
existence of a wealth of strange and non-strange s- and d-wave wave baryon resonances, once the
crossing symmetry constraint is implemented properly.
Naively one may expect that chiral dynamics does not make firm
predictions for d-wave resonances since the meson baryon-octet
interaction in the relevant channels probes a set of
counter terms presently unknown. However, this is not
necessarily so. Since a d-wave baryon resonance couples to s-wave
meson baryon-decuplet states chiral symmetry is quite predictive
for such resonances under the assumption that the
latter channels are dominant. This is in full analogy to the
analysis of the s-wave resonances  \cite{Wyld,Dalitz,Ball,Rajasekaran,Wyld2,Granada,Ji01,Jido03}
that neglects the effect of the contribution of d-wave meson baryon-decuplet
states. The empirical observation that the d-wave resonances
$N(1520)$, $N(1700)$ and $\Delta (1700)$ have large branching
fractions ($> 50 \% $) into the inelastic $N \pi \pi$ channel, even
though the elastic $\pi N$ channel is favored by phase space,
supports our assumption.

\section{Effective field theory of chiral coupled-channel dynamics}

The starting point to describe the meson-baryon scattering process
is the chiral SU(3) Lagrangian (see e.g.\cite{Krause}). A systematic approximation
scheme arises due to a successful scale separation justifying the chiral
power counting rules \cite{book:Weinberg}. Our effective field theory of the meson-baryon
scattering processes is based on the assumption that the scattering amplitudes are
perturbative at subthreshold energies with the expansion parameter
$Q/ \Lambda_{\chi}$. The small scale $Q$ is to be identified with any small
momentum of the system. The chiral symmetry breaking scale is
$$\Lambda_\chi \simeq 4\pi f \simeq 1.13 \;{\rm GeV}\,, $$ with the parameter $f\simeq 90$ MeV
determined by the pion decay process. Once the available
energy is sufficiently high to permit elastic two-body scattering a further typical
dimensionless parameter $m_K^2/(8\,\pi f^2) \sim 1$ arises \cite{LK00,LK01,LK02}. Since this
ratio is uniquely linked to two-particle reducible diagrams it is sufficient to sum
those diagrams keeping the perturbative expansion of all irreducible
diagrams, i.e. the  coupled-channel Bethe-Salpeter equation has to be solved. This is the
basis of the $\chi$-BS(3) approach developed in \cite{LK00,LK01,LK02,grnpi,grkl}.

At leading order in the chiral expansion one encounters the Weinberg-Tomozawa \cite{Wein-Tomo} interaction,
\begin{eqnarray}
&&\mathcal{L}_{WT}=
 \frac{i}{8\, f^2}\, {\rm tr}\, \Big((\bar B
\,\gamma^\mu\,\, B) \cdot
 [\Phi,(\prt_\mu \Phi)]_-  \Big)
+\frac{3\,i}{8\, f^2}\, {\rm tr}\, \Big((\bar B_\nu
\,\gamma^\mu\,\, B^\nu) \cdot
 [\Phi,(\prt_\mu \Phi)]_-  \Big)
 \,,
 \label{WT-term}
\end{eqnarray}
where we dropped additional structures that do not contribute to the on-shell scattering process
at tree level. The terms in (\ref{WT-term}) constitute the leading order s-wave interaction
of Goldstone bosons ($\Phi $) with the baryon-octet ($B$) and baryon-decuplet ($B_\mu$) states.
The octet and decuplet fields, $\Phi, B$ and $B_\mu$, posses an appropriate matrix structure
according to their SU(3) tensor representation.

The scattering process is described by the amplitudes that follow as solutions of the
Bethe-Salpeter equation,
\begin{eqnarray}
&& T(\bar k ,k ;w ) = V(\bar k ,k ;w )
+\int\!\! \frac{d^4l}{(2\pi)4}\,V(\bar k ,
l;w )\, G(l;w)\,T(l,k;w )\;,
\nonumber\\
&& G(l;w)=-i\,D({\textstyle{1\over 2}}\,w-l)\,S( {\textstyle{1\over 2}}\,w+l)\,,
\label{BS-coupled}
\end{eqnarray}
where we suppress the coupled-channel structure for simplicity. The meson and
baryon propagators,  $D(q)$ and $S(p)$, are used in the notation of \cite{LWF02}.
We apply the convenient kinematics:
\begin{eqnarray}
&& w = p+q = \bar p+\bar q\,,
\quad k= {\textstyle{1\over 2}}\,(p-q)\,,\quad
\bar k ={\textstyle{1\over 2}}\,(\bar p-\bar q)\,,
\label{def-moment}
\end{eqnarray}
where $q,\,p,\, \bar q, \,\bar p$ the initial and final meson and baryon 4-momenta.
The Bethe-Salpeter scattering equation is recalled for the case of meson baryon-octet
scattering. An analogous equation holds for meson baryon-decuplet scattering process
(see e.g. \cite{LWF02}). Referring to the detailed discussion given in \cite{LK02,LK03,Lutz00,NA00} we assume a
systematic on-shell reduction of the Bethe-Salpeter interaction kernel
leading to the effective interaction $V$ used in (\ref{BS-coupled}).
The latter is is expanded according to chiral power counting rules.
The scattering amplitude $T(\bar k,k;w)$ decouples into various sectors
characterized by isospin ($I$) and strangeness ($S$) quantum numbers. In the
case of meson baryon-octet and baryon-decuplet scattering the following channels are relevant
\begin{eqnarray}
&& (I,S)_{[8 \otimes 8]} =  (0,-3), (1,-3), (\frac{1}{2},-2), (\frac{3}{2},-2) ,
(0,-1),  \nonumber\\
&& \qquad \qquad (1,-1),
(2,-1), (\frac{1}{2},0),(\frac{3}{2},0), (0,1),(1,1) \,,
\nonumber\\
&& (I,S)_{[8 \otimes 10]} = (\frac{1}{2},-4), (0,-3), (1,-3), (\frac{1}{2},-2), (\frac{3}{2},-2) ,
(0,-1), \nonumber\\
&& \qquad \qquad  (1,-1),
(2,-1),(\frac{1}{2},0),(\frac{3}{2},0), (\frac{5}{2},0), (1,1),(2,1) \,.
\label{sectors-10}
\end{eqnarray}
Following the $\chi$-BS(3) approach developed in \cite{LK02,LWF02} the effective
interaction kernel is decomposed into a set of covariant projectors that have well
defined total angular momentum, $J$, and parity, $P$,
\begin{eqnarray}
&& V(\bar k ,k ;w )  = \sum_{J,P}\,V^{(J,P)}(\sqrt{s}\,)\,
{\mathcal Y}^{(J,P)}(\bar q, q,w) \,.
\label{def-proj}
\end{eqnarray}
The merit of the projectors is that they decouple the Bethe-Salpeter
equation (\ref{BS-coupled}) into orthogonal sectors labelled by the total
angular momentum, $J$, and parity, $P$. We insist on the renormalization condition,
\begin{eqnarray}
&&T^{(I,S)}(\bar k,k;w)\Big|_{\sqrt{s}= \mu (I,S)} =
V^{(I,S)}(\bar k,k;w)\Big|_{\sqrt{s}= \mu (I,S)} \,,
\label{ren-cond}
\end{eqnarray}
together with the natural choice for the subtraction points,
\begin{eqnarray}
&& \mu(I,+1)=\mu(I,-3)={\textstyle{1\over 2}}\,(m_\Lambda+ m_\Sigma) \,,
\quad \mu(I,0)=m_N\,, \quad
\nonumber\\
&&  \mu(0,-1)=m_\Lambda,\quad \mu(1,-1)=m_\Sigma\,, \quad
\mu(I,-2)= \mu(I,-4)= m_\Xi
 \label{eq:sub-choice}
\end{eqnarray}
as explained in detail in \cite{LK02}. The renormalization condition reflects
the crucial assumption our effective field theory is based on, namely that
at subthreshold energies the scattering amplitudes can be
evaluated in standard chiral perturbation theory. This is achieved
by supplementing (\ref{BS-coupled}) with (\ref{ren-cond},\ref{eq:sub-choice}).  The
subtraction points
(\ref{eq:sub-choice}) are the unique choices that protect the s-channel
baryon-octet masses manifestly in the p-wave $J={\textstyle{1\over 2}}$
scattering amplitudes. The merit of the scheme \cite{LK00,LK01,LK02} lies in the
property that for instance the $K \,\Xi$ and $\bar K\,\Xi $
scattering amplitudes match at $\sqrt{s} \sim m_\Xi $
approximately as expected from crossing symmetry. In \cite{LK02} we suggested
to glue s- and u-channel unitarized scattering amplitudes at subthreshold energies.
This construction reflects our basic assumption that diagrams showing an s-channel
or u-channel unitarity cut need to be summed to all orders at least at energies close to
where the diagrams develop their imaginary part. By construction, a glued scattering amplitude
satisfies crossing symmetry exactly at energies where the scattering process takes
place. At subthreshold energies crossing symmetry is implemented
approximatively only, however, to higher and higher accuracy when more chiral correction
terms are considered. Insisting on the renormalization condition
(\ref{ren-cond},\ref{eq:sub-choice}) guarantees that subthreshold amplitudes match smoothly
and therefore the final 'glued' amplitudes comply with the crossing-symmetry constraint
to high accuracy. The natural subtraction points
(\ref{eq:sub-choice}) can also be derived if one incorporates photon-baryon
inelastic channels. Then additional constraints arise. For instance the
reaction $\gamma \,\Xi \to \gamma \,\Xi $,
which is subject to a crossing symmetry constraint at threshold, may
go via the intermediate states $\bar K \,\Lambda $ or $\bar K \,\Sigma $.

The perturbative nature of subthreshold amplitudes, a crucial
assumption of the $\chi$-BS(3) approach proposed in \cite{LK00,LK01,LK02}, is
not necessarily true in phenomenological coupled-channel schemes
in \cite{ksw95,grnpi,grkl,Oset-prl,Oset-plb,Jido03}. Using the subtraction scales
as free parameters, as advocated in \cite{Oset-prl,Oset-plb,Jido03}, may be viewed as
promoting the counter terms of chiral order $Q^3$ to be unnaturally large. If the
subtraction scales are chosen far away from their natural values (\ref{eq:sub-choice})
the resulting loop functions are in conflict with chiral power counting rules \cite{LK00}.
Though unnaturally large $Q^3$ counter terms can not be excluded from first principals
one should check such an assumption by studying corrections terms systematically \cite{LK01,LK02}.
A detailed test of the naturalness of the $Q^3$ counter terms was performed within the $\chi$-BS(3)
scheme \cite{LK02} demonstrating good convergence in the channels studied
without any need for promoting the counter terms of order $Q^3$. Possible
correction terms in the approach followed in \cite{Oset-prl,Oset-plb,Jido03}
have so far not been studied systematically for meson-baryon scattering. Moreover,
if the scheme advocated in \cite{Oset-prl,Oset-plb,Jido03} were applied in all eleven isospin
strangeness sectors with $J^P=\frac{1}{2}^-$  a total number of 26
subtraction parameters arise. This should be compared with the only ten counter terms of chiral
order $Q^3$ contributing to the on-shell scattering amplitude at that order \cite{LK02}.
Selecting only the operators that are leading in the large-$N_c$ limit of QCD out of the
ten $Q^3$ operators only four survive \cite{LK02}. 
For consistency and before applying the approach of \cite{Oset-prl,Oset-plb,Jido03}
to all isospin strangeness channels, it would be highly desirable to
address the above mismatch of parameters. Our scheme has the advantage that once the parameters
describing subleading effects are determined in a subset of sectors one has immediate
predictions for all sectors $(I,S)$. A mismatch of the number of parameters is avoided
altogether since the  $Q^3$ counter terms enter the effective interaction kernel directly.

Given the subtraction scales (\ref{eq:sub-choice}) the leading order calculation is
parameter free. Of course chiral correction terms do lead to further so far unknown
parameters  which need to be adjusted to data. Within the $\chi-$BS(3) approach such
correction terms enter the effective interaction kernel $V$ rather than leading to
subtraction scales different from (\ref{eq:sub-choice}) as it is assumed
in~\cite{Oset-prl,Oset-plb,Jido03}. In particular the leading
correction effects are determined by the counter terms of chiral
order $Q^2$. The effect of altering the subtraction scales away
from their optimal values (\ref{eq:sub-choice}) can be compensated
for by incorporating counter terms in the chiral Lagrangian that
carry order $Q^3$.

\section{Baryon resonances from chiral SU(3) symmetry}

We begin with a discussion of the s- and d-wave baryon resonance spectrum that arises in
the SU(3) limit. The latter is not defined uniquely depending on the magnitude of the current
quark masses, $m_u=m_d=m_s$. We consider two scenarios \cite{Granada,KL03}. In the 'light' SU(3)
limit the current quark masses are chosen such that one obtains $m_\pi= m_K=m_\eta =140$ MeV.
The second case, the 'heavy' SU(3) limit, is characterized by $m_\pi= m_K=m_\eta =500$ MeV.
In the SU(3) limit meson baryon-octet scattering is classified according to,
\begin{eqnarray}
&& 8 \otimes 8 = 27 \oplus \overline{10} \oplus 10 \oplus 8 \oplus 8 \oplus 1 \,.
\label{8-8-decom}
\end{eqnarray}
The leading order chiral Lagrangian predicts attraction in the
two octet and the singlet channels but repulsion in the 27-plet and decuplet channels.
As a consequence in the 'heavy' SU(3) limit the chiral dynamics predicts two degenerate
octet bound states together with a non-degenerate singlet state
\cite{Wyld,Dalitz,Ball,Rajasekaran,Wyld2,Granada,Jido03,Granada,KL03}.
In the 'light' SU(3) limit all states disappear.
In the $J^P=\frac{3}{2}^-$ sector the Weinberg-Tomozawa interaction is attractive in
the octet, decuplet and 27-plet channels, but repulsive in the 35-plet channel,
\begin{eqnarray}
&&8 \otimes 10 = 35 \oplus 27 \oplus 10 \oplus 8\,.
\label{8-10-decom}
\end{eqnarray}
Therefore one may expect resonances or bound states in the former channels.
Indeed, in the 'heavy' SU(3) limit we find $72=4\times (8+10)$ bound states in this sector
forming an octet and decuplet representation of the SU(3) group. We do not find a
27-plet-bound state reflecting the weaker attraction in that channel. However, if we
artificially increase the amount of attraction by about 40 $\%$ by lowering the value of $f$
in the Weinberg-Tomozawa term, a clear bound state arises in this channel also.
A contrasted result is obtained if we lower the meson masses down to the pion mass arriving at
the 'light' SU(3) limit. Then we find neither bound nor resonance octet or decuplet states.
This pattern is a clear prediction of chiral coupled-channel dynamics
which should be tested with unquenched QCD lattice simulations \cite{dgr}.

\begin{figure}
\resizebox{0.5\textwidth}{!}{%
  \includegraphics{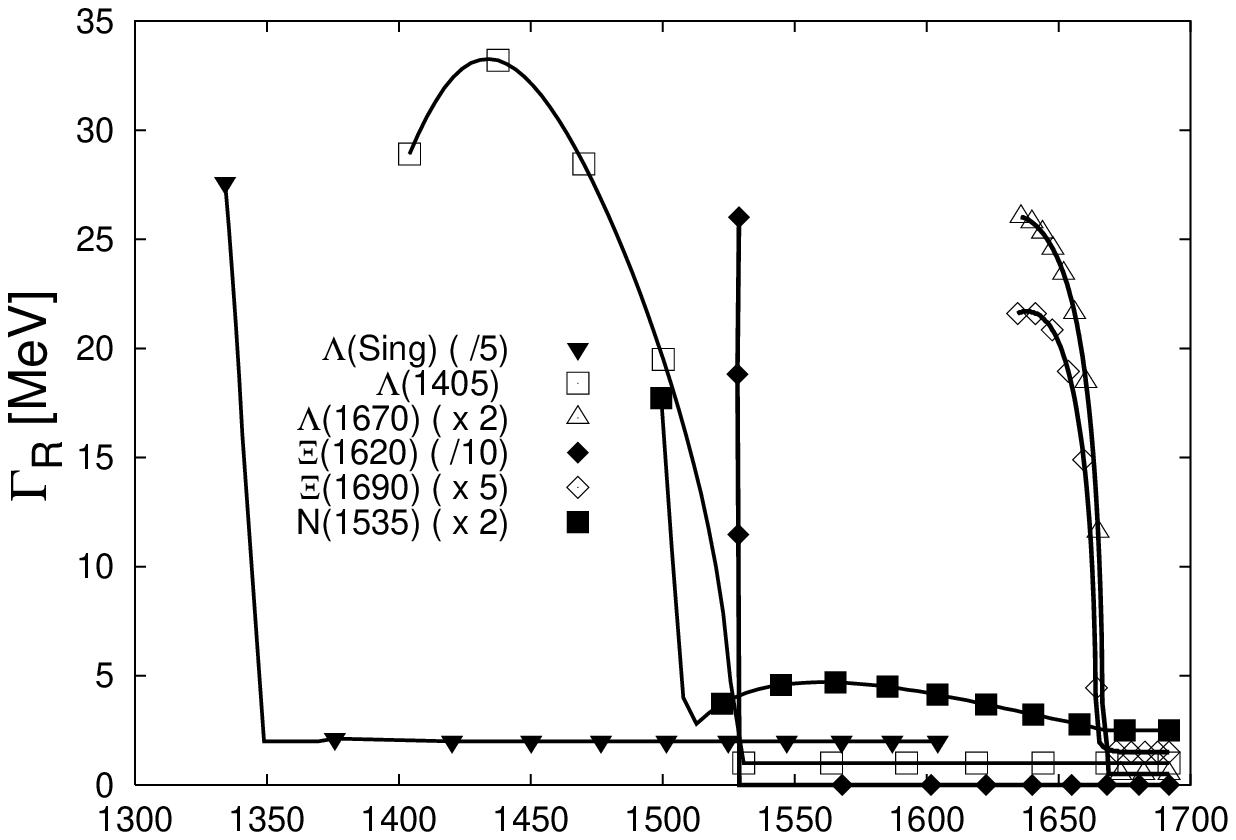}}
\resizebox{0.5\textwidth}{!}{%
  \includegraphics{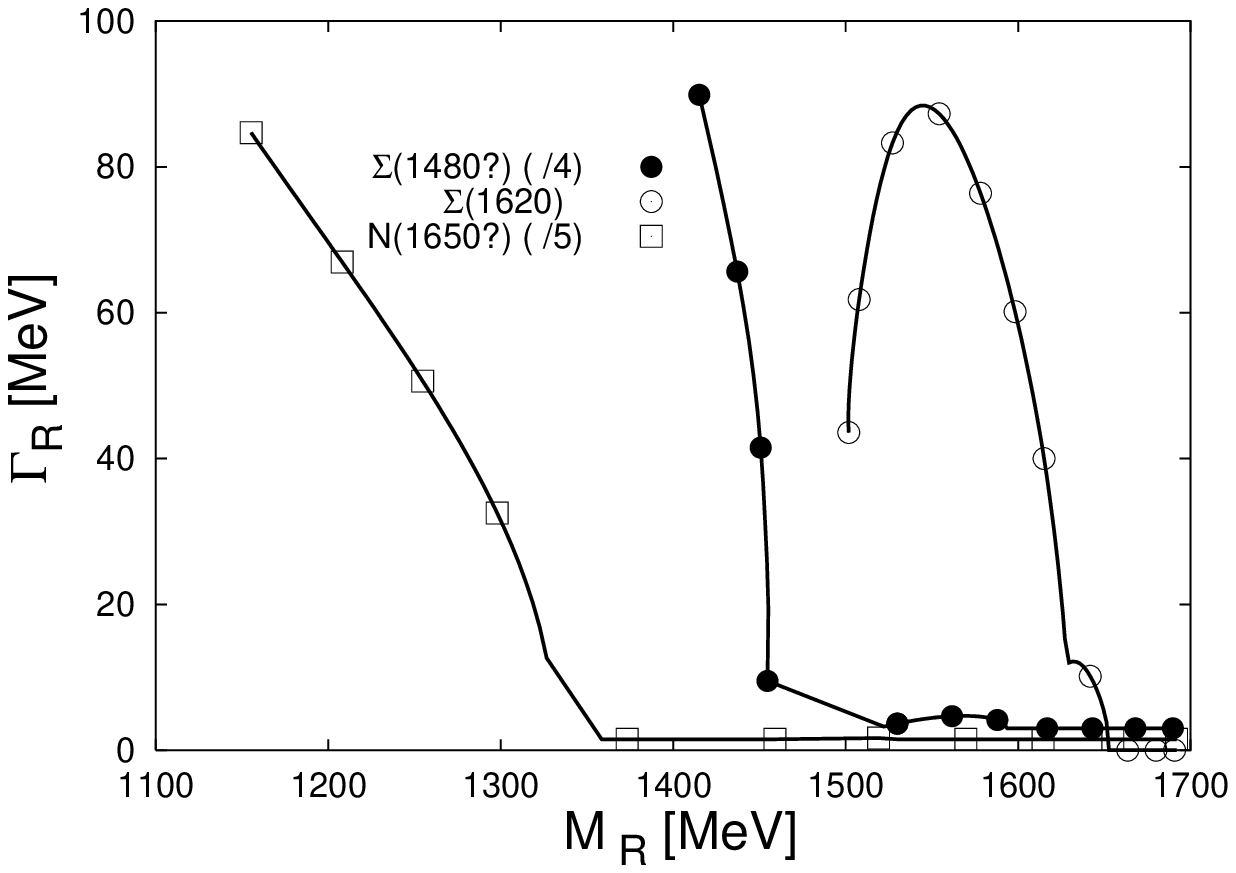}}
\vskip-0.9cm
\caption{\footnotesize Masses and decay widths of two octets and a
singlet of baryon states for several values of the pion mass. For each
baryon state we plot eleven points, which correspond to eleven equally spaced
values of $x$ (see~(\protect\ref{eq:su(3)})) ranging from 1 (first point
from the right) to 0 (first point from the left). Widths of different
baryon resonances have been scaled by factors, as it is  detailed in
the legend of the plots. To disentangle among different states, for
some of them, the widths have been shifted by constants
factors. Lines have been plotted just to guide the eye.}
\label{fig:1}       
\end{figure}

We look for poles on the Second Riemann Sheet of our s-wave scattering
amplitudes. Positions of the poles determine masses and widths of the resonances
while the residues in the different channels define the
branching ratios ~\cite{grnpi,grkl}.
To explore the quark mass dependence of the resonances, we vary
the averaged $up$ and $down$ quark masses, but keep fixed the
antikaon mass. A parameter $x$ is introduced in
terms of which the pion mass varies as
\begin{eqnarray}
&& \left. m_\pi^2\right|_{\rm SU(3)} = m_\pi^2 + x (m_K^2-m_\pi^2),
\quad x\in[0,1] \,.
\label{eq:su(3)}
\end{eqnarray}
A pair $(m_K^2,
\left. m_\pi^2\right|_{\rm SU(3)})$ determines the $\eta$ meson mass
via the Gell-Mann--Oakes--Renner relation. Given the SU(3) symmetry breaking
parameters $b_0$, $b_D$ and $b_F$ of \cite{LK02}, the masses of the
baryon octet (N(940),\,$\Lambda(1115)$,\,$\Sigma(1190)$ and $\Xi(1320)$)
are also determined.  In the limit $x=1$ our SU(3) pion is as heavy as the
real kaon, while when $x=0$ the physical world is recovered approximatively.
As demonstrated in Fig. \ref{fig:1} we find a remarkable success
predicting rather well the bulk of the features of the four stars
N(1535), $\Lambda (1405)$ and $\Lambda (1670)$ resonances \cite{Granada}.

We also find a narrow  resonance with $(I,S)=(1,-1)$, though it appears in the unphysical
11000 sheet (see \cite{grnpi,Granada}). In Fig. \ref{fig:1} the trajectory of this pole is labelled with
$\Sigma(1620)$, a two star resonance, since it couples strongly to the $\pi\Lambda$, $\pi\Sigma$ and specially to the
${\bar K}{\rm N}$ channels. There exists also a pole
in the 11100 sheet with $M_R=1466$ MeV and $\Gamma_R=574$ MeV not shown in Fig. \ref{fig:1}.
It is above the ${\bar K}{\rm N}$ threshold and has a large coupling to the $K\Xi$ channel. This
very broad pole is precisely the one quoted in \cite{Ji01}. It does not influence the physical
scattering at all. However, chiral corrections reduce its width significantly and take its mass closer to the $K\Xi$
threshold indicating that it is the three star $\Sigma(1750)$ resonance~\cite{LK02}.
Fig. \ref{fig:1} shows also the fate of the second $(I,S)=(\frac{1}{2},0)$ state, which is moved far away from the
real axis down to quite low energies. Chiral corrections appear not to
take the position of this pole closer to that of the physical N(1650) resonance \cite{LK02}. A quantitative
description may require the inclusion of further inelastic channels,
like $\pi \Delta$ and $\rho N$ \cite{LWF02}

In the $S=-2$ sector we find two
resonances, which can clearly be identified to the $\Xi(1690)$ and
$\Xi(1620)$ resonances. Of particular interest is the signal for the
$\Xi(1690)$ resonance, where we find a quite small (large) coupling to
the $\pi\Xi$ (${\bar K}\Sigma$) channel, which explains the smallness
of the experimental ratio, $\Gamma(\pi \Xi) / \Gamma ({\bar K} \Sigma)
< 0.09 $~\cite{pdg} despite of the significant energy difference between the
thresholds for the $\pi \Xi$ and $ {\bar K} \Sigma$ channels. On the other
hand, we also find a third $\Lambda$ resonance placed also below the ${\bar K} {\rm N}$ threshold and with a
large coupling to the $\pi \Sigma$ channel. This confirms the findings
of the recent works \cite{Jido03,Ji01,grkl}, where the dynamics of
the $\Lambda(1405)$ states was studied.

\begin{figure}[t]
\begin{center}
\includegraphics[width=11.7cm,clip=true]{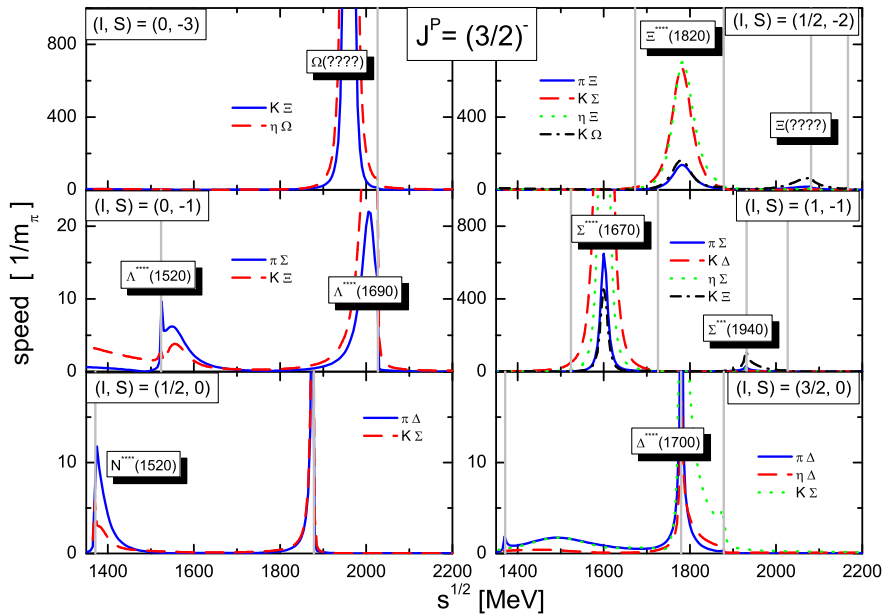}
\end{center}
\vskip-1.2cm
\caption{\footnotesize Diagonal speed plots of the $J^P=\frac{3}{2}^-$ sector. The vertical lines show the
positions of inelastic meson baryon-decuplet channels. Parameter-free results are obtained
in terms of physical masses and $f=90$ MeV.} \label{fig1}
\end{figure}

In Fig. \ref{fig1} speed plots of the $J^P=\frac{3}{2}^-$
sector are shown for all channels in which octet and decuplet resonance states are expected.
In \cite{KL03} we generalized the notion of a
speed \cite{Hoehler:speed} to the case of coupled-channels in way that the latter reveals the coupling strength
of a given resonance to any channel, closed or open. If a resonance with not too large decay
width sits in the amplitude a clear peak structure emerges in the speed plot even if the
resonance structure is masked by a background phase. In the case of s-wave scattering thresholds
induce square-root singularities which should not be confused with a resonance signal.
It is a remarkable success of the $\chi$-BS(3) approach that it predicts the
four star hyperon resonances $\Xi(1820)$, $\Lambda(1520)$, $\Sigma (1670)$ with masses quite
close to the empirical values. The nucleon and isobar resonances $N(1520)$ and $\Delta (1700)$
also present in Fig. \ref{fig1}, are predicted with less accuracy. The important result here is the
fact that those resonances are generated at all. It should not be expected to obtain already
fully realistic results in this leading-order calculation. For instance chiral correction
terms are expected to provide a d-wave $\pi \,\Delta$-component of the $N(1520)$.
We continue with the peak in the (0,-3)-speeds at mass 1950 MeV.
Since this is below all thresholds it is in fact a bound state. Such a state
has so far not been observed but is associated with a decuplet resonance \cite{Schat}.
Further states belonging to the decuplet are seen in the $(\frac{1}{2},-2)$- and
$(1,-1)$-speeds at masses 2100 MeV and 1920 MeV. The latter state can be identified with the
three-star $\Xi (1940)$ resonance. Finally we point at the fact that the $(0,-1)$-speeds show
signals of two resonance states consistent with the existence of the four star resonance
$\Lambda(1520)$ and $\Lambda(1690)$ even though in the 'heavy' SU(3) limit we observed only
one bound state. It appears that the SU(3) symmetry breaking pattern generates the 'missing'
state in this particular sector by promoting the weak attraction of the 27-plet contribution
in (\ref{8-10-decom}).

\section{Summary}

In this talk we reported on recent progress in the understanding of baryon resonances
based on chiral-coupled channel dynamics. The reader was introduced to
an effective field theory formulation of chiral coupled-channel dynamics. Leading order
results predict the existence of s- and d-wave baryon resonances with a spectrum remarkably
close to the empirical pattern without any adjustable parameters. The formation of
resonances is a consequence of the chiral SU(3) symmetry of QCD, i.e. in an effective field
theory, that was based on the chiral SU(2) symmetry only, no resonances would be formed.


\begin{thebibliography}{9}

\bibitem{LK01}
M.F.M. Lutz and E.E. Kolomeitsev, Found. Phys. {\bf 31} (2001) 1671.

\bibitem{LK02} M.F.M. Lutz and E. E. Kolomeitsev,
Nucl. Phys. {\bf A 700} (2002) 193.

\bibitem{LWF02}
M.F.M. Lutz, Gy. Wolf and B. Friman, Nucl. Phys. {\bf A 706} (2002) 431.

\bibitem{LH02}
M.F.M. Lutz, GSI-Habil-2002-1.

\bibitem{Wyld} H.W. Wyld, Phys. Rev. {\bf 155} (1967) 1649.

\bibitem{Dalitz} R.H. Dalitz, T.C. Wong and G. Rajasekaran,
Phys. Rev. {\bf 153} (1967) 1617.

\bibitem{Ball}
J.S. Ball and W.R. Frazer, Phys. Rev. Lett. {\bf 7} (1961) 204.

\bibitem{Rajasekaran}
G. Rajasekaran, Phys. Rev. {\bf 5} (1972) 610.

\bibitem{Wyld2}
R.K. Logan and H.W. Wyld, Phys. Rev. {\bf 158} (1967) 1467.

\bibitem{sw88} P.B.~Siegel and W.~Weise,
Phys.\ Rev.\ C {\bf 38} (1988) 2221.

\bibitem{Wein-Tomo} S. Weinberg, Phys. Rev. Lett. {\bf 17} (1966) 616;\\
Y. Tomozawa, Nuov. Cim. {\bf A 46} (1966) 707.

\bibitem{LK00}
M.F.M. Lutz and E. E. Kolomeitsev, Proc. of Int. Workshop XXVIII
on Gross Properties of Nuclei and Nuclear Excitations, Hirschegg,
Austria, January 16-22, 2000.

\bibitem{Granada} 
C. Garc\'\i a-Recio, M.F.M. Lutz and J. Nieves, 
nucl-th/0305100, Phys. Lett. B in print.

\bibitem{KL03}
E.E. Kolomeitsev and M.F.M. Lutz, 
nucl-th/0305101, Phys. Lett. B in print.

\bibitem{Ji01} D. Jido, et al.,
Phys. Rev. {\bf C 66} (2002) 055203.

\bibitem{Jido03}
D. Jido et al., Nucl.\ Phys.\ {\bf A 725} (2003) 181.

\bibitem{Krause}
A.~Krause, Helv. Phys. Acta {\bf 63} (1990) 3.

\bibitem{book:Weinberg}
S. Weinberg, {\it The quantum theory of fields}, Vol. II, University Press, Cambridge (1996).

\bibitem{grnpi} J. Nieves and E. Ruiz Arriola, Phys. Rev. {\bf
D 63}, (2001) 076001.

\bibitem{grkl} C. Garc\'\i a-Recio et al., Phys. Rev. {\bf D 67} (2003) 076009.

\bibitem{LK03}
M.F.M. Lutz and E.E. Kolomeitsev, nucl-th/0307039, Nucl. Phys. A in print.

\bibitem{Lutz00}
M.F.M. Lutz, Nucl. Phys. {\bf 677} (2000) 241.

\bibitem{NA00}
J. Nieves and E. Ruiz Arriola, Nucl. Phys. {\bf A 679} (2000) 57.

\bibitem{ksw95} N.~Kaiser, P.B.~Siegel and W.~Weise,
Nucl.\ Phys.\ A {\bf 594} (1995) 325.

\bibitem{Oset-prl} A. Ramos, E. Oset and C. Bennhold,
Phys. Rev. Lett. {\bf 89} (2002) 252001.

\bibitem{Oset-plb}
E. Oset, A. Ramos, C. Bennhold,  Phys. Lett. {\bf B 527} (2002)
99.

\bibitem{dgr} D. G. Richards, Proc. of `NSTAR 2002', Pittsburgh,
October 2002, and references therein.

\bibitem{pdg} K. Hagiwara et al., Phys. Rev. {\bf D 66} (2002) 010001.

\bibitem{Hoehler:speed}
G. H\"ohler, $\pi N$ NewsLetter {\bf 9} (1993) 1.

\bibitem{Schat}
C.L. Schat, J.L. Goity and N.N. Scoccola, Phys. Rev. Lett. {\bf
88} (2002) 102002.

\end{thebibliography}
\end{document}